\documentclass[12pt,preprint]{aastex}

\newcommand{\beq}{\begin{equation}}
\newcommand{\eeq}{\end{equation}}

\begin{document}

\markboth{Zhou \& Sun} {Transition to chaos in N-planet systems}

%
%

\title{Do N-planet systems have a boundary between chaotic and regular motions?}

\author{Ji-Lin Zhou ,  Yi-Sui SUN}

\affil{Department of Astronomy, Nanjing University, Nanjing
210093, China; zhoujl@nju.edu.cn}



\begin{abstract}
 Planetary systems consisting of one star and $n$ planets
with equal planet masses $\mu$ and scaled orbital separation
are referred as EMS systems.
 They represent an  ideal model for planetary systems during the post-oligarchic evolution.
Through the calculation of Lyapunov exponents, we study the boundary between
chaotic and regular regions of EMS systems. We find that for $n \ge 3$,
there does not exist a transition region in the  initial separation space,
 whereas  for $n=2$, a clear borderline  occurs
with relative separation $\sim  \mu^{2/7}$ due to overlap of
resonances (Wisdom, 1980).
This phenomenon is caused by the slow diffusion of
velocity dispersion ($\sim t^{1/2}$, $t$ is the time) in  planetary systems
with $n \ge 3$, which leads to chaotic motions at the time of roughly two orders of magnitude before
the orbital crossing occurs.
 This result does not conflict with
the existence of transition boundary in the full phase space of  N-body systems.

\end{abstract}

\keywords{N-body Problem; Stability; Chaos; Diffusion.}

\section{Introduction}

Planetary system is a paradigm of classical N-body problems.
The pioneer work of Poincar{\'e}\cite{1} indicated
that a general  N-body system with $N \ge 3$ is  not integrable.
According to the work of Arnold\cite{2}, a typical near-integrable Hamiltonian system
with more than $3$ degrees of freedom is topologically unstable, i.e, the action variables can
diffuse away in the phase space  during a sufficient long period.
 However, for a specific planetary system, when
the instability is significant and how long the system can be stable
are unsolved problems with great astronomical interests\cite{3}.

 One of the most interesting questions is whether there exists a
 transition region between
regular and chaotic motions in a $n$-planet system.
 Gladman\cite{4} found that for $n=2$,
 the orbits of two planets with equal masses ($\mu$) will be chaotic as
 long as their relative separation
 $\ge 2 \mu^{2/7}$. The scaling $\mu^{2/7}$ was first found by Wisdom\cite{5} in
 a restricted three-body model with a resonance overlap criterion.
 However, whether there exist similar relationships for systems with $n\ge 3$   remains unknown.

In a previous paper(\cite{6}, hereafter Paper I), we investigated  the orbital crossing time and
orbital diffusion of planetary systems  with equal planet masses
and scaled separation (we call them the EMS systems).
 We found that for EMS systems with $n\ge 3$, the velocity dispersion of
the planets diffuses with time $t$ as $\sim t^{1/2}$.
In this paper, we study
the transition from regular to chaotic regions in the initial separation space
of these systems.

\section{Model and Method}

An  EMS system consists of a star with unit mass ($\mu_0=1$) and
$n$ planets  with equal masses $\mu~ (\mu \ll 1)$ moving around it.
 Assume  all the planets are initially in circular and coplanar orbits
 with semi-major axes $a_i $, where $i=1,...,n$ are sequenced from inner to outside.
  We put the innermost planet at $a_1=1$AU, which is the average distance
  from the Sun to the Earth. The  initial separation
  between the planetary orbits, denoted as $k_0$,
   are  equal when scaled by their mutual Hill's radii $R_H$,
\beq
 k_0=\frac{ a_{i+1}-a_i}{R_H},~~ R_H= (\frac{2\mu}{3} )^{1/3}\frac{a_i+a_{i+1}}{2},~~(i=1,...,n-1).
\label{kbe}
\eeq
The initial mean anomaly $M_i$ and longitude
of perihelion $\varpi_i$ of each planet orbit are chosen randomly.

In an inertial coordinate with origin on the mass center of
the $n+1$ bodies ( the star and $n$ planets), the Hamiltonian of the system is given as,
 \beq
H=\sum_{i=0}^{n} \frac{{\bf p}_i^{2}}{2\mu_i} -\sum_{i=0}^{n}
\sum_{j=0,j>i}^{n} \frac{G\mu_i\mu_j}{r_{ij}},
\label{ham}
\eeq
where $ {\bf r}_i$ and ${\bf p_i}= \mu_i {\dot {\bf r}_i}~(i=0,...n)$
are the position and momentum of the star and the planets, respectively,
$r_{ij}=|{\bf r}_j-{\bf r}_i|$ is the distance between bodies $i$ and $j$, G is the gravitational constant.

The  equations of motion corresponding to eq. (\ref{ham})
are integrated by a Runge-Kutta method with adaptive
step-sizes. The orbits are simulated up to $10^7$ years, with an accumulated error of
relative energy $\sim 10^{-8}$ for weak chaotic or regular orbits and
$\sim 10^{-6}$ for strong chaotic orbits in the case of $n=3$ (Fig.1a).
 The Lyapunov exponents are defined as the limit of
$\chi(t)$ when time $t$ tends to infinity, with
\beq
 \chi(t) = \frac{1}{t}\log (\frac{|\xi(t)|}{|\xi(0)|}),
\eeq
where $\xi(t)$ and $\xi(0)$ are tangential vectors in the phase space at time  $t$
and 0, respectively, $|\cdot|$ is the norm of a vector.
 We solve the variational equations of eq. (2) along the solution to get $\chi(t)$.
Normalization of $\xi(t)$  at every fixed time interval is required
to avoid overflow during the simulations.

\section{Numerical Results}

\begin{figure}[bt]
\includegraphics{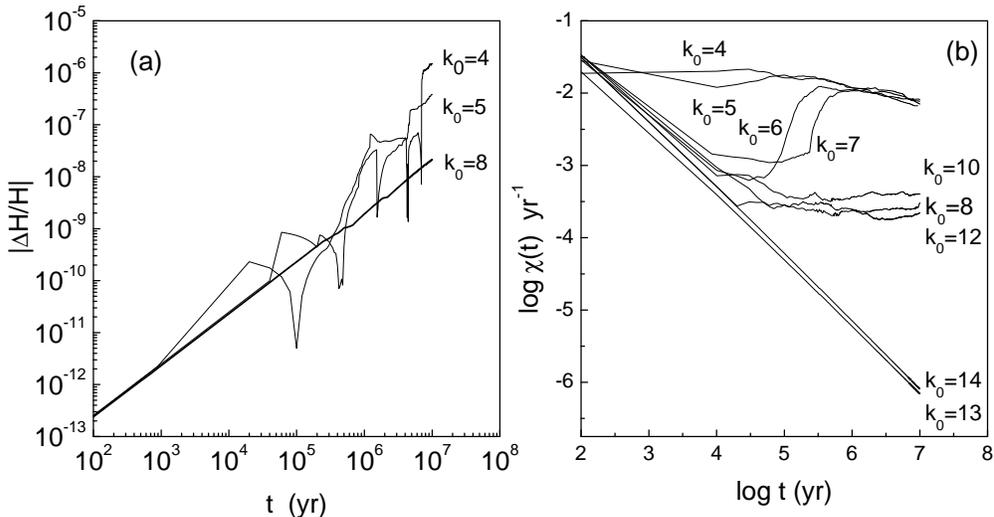}
\vspace{6cm}
\caption{ Evolution of 3-plant EMS systems with mass $\mu=10^{-8}$.
 (a) Variations of accumulated error of the relative energy with time. The three
lines correspond to initial separation  $k_0=4,5,8$, respectively.   (b)
Variations of $\chi(t)$ for orbits with different initial separation $k_0$. }
\end{figure}

A notable phenomenon  in an EMS system with $n\ge 3$
is for some orbits,  $\xi(t)$ may reach a limit of medium  value ( $\sim 10^{-3}~{\rm yr}^{-1}$)
at time $t\sim 10^5$ yrs, then it  grows to  a large limit ( $\sim 10^{-2}~{\rm yr}^{-1}$ )
 after a long period time ( $\sim 10^7$ yrs),
 indicating that the orbits finally become quite chaotic
after a long time in a transitional state (e.g., orbits with $k_0=6,7$ in Fig.1b).
Moreover, we find that these transition orbits are in the
border region between strong (with large Laypunov exponents) and weak (with medium ones)
chaotic motions.

The transition phenomenon of $\chi(t)$  does not
occur in EMS systems with $n=2$. As shown in Fig.2a, different
integration time spans give almost identical border $k_c$ between  chaotic and
regular motions. For $\mu=10^{-8}$, $k_c\approx 5.2$,
which roughly
coincides with the location predicted by Wisdom and Gladman's criterion, i.e.,  $k_c=2\mu^{2/7}/(2\mu/3)^{1/3}\approx 2.3 \mu^{-1/21}
 \approx 5.5$.

The evolution of $\chi(t)$ is linked with the velocity
dispersion ($\sigma$) in the planetary systems.
  As demonstrated in paper I, one of the major characteristics  of planetary systems
  with $n\ge 3$ is that $\sigma$ diffuses slowly with $\sim t^{1/2}$
    due to the interactions between planets.
The growth of velocity dispersion  (orbital eccentricities) of the
planets enhances the planetary perturbations,
leads to the growth of $\chi(t)$ after a transition stage, and finally
 orbital crossing may occur after the systems undergo a long period of strong chaotic motions.
Define the orbital crossing time $T_c$ as
the minimum duration that either orbit crossing ( $a_i \ge  a_{j}$ for $i<j$)
or close encounter ($r_{ij} < R_H$) occurs.
In paper I, we  derived an empirical formula of
$T_c $ for $ k_0>2.3, 10^{-4}\le \mu \le 10^{-10}$. For planets in circular orbits, it gives,
\beq
 \log (\frac{T_{\rm c}}{\rm yr}) =(-2-0.27\log\mu) + (18.7+ 1.1\log\mu) \log(\frac{k_0}{2.3}) .
 \label{ff}
\eeq
Fig.2b plots the $\chi(10^7 {\rm yr})$ and quantities $-\log(T_c)$. A relationship
 $\log(\chi(10^7 {\rm yr}))\sim 2- \log(T_c)$ suggests that the time scale that
an orbital becomes chaotic (i.e. Lyapunov time ) is roughly two orders of magnitude
less than the orbital crossing time. The correlation indicates that the
growth of $\chi(t)$ is due to the slow orbital diffusion towards
 high velocity dispersion.
Since growth of velocity dispersion is a robust procedure in a planetary system with $n\ge 3$,
we  believe that the transitional regular orbits will eventually become chaotic after sufficient
long time, hence there is no clear boundary between the domains of chaotic
and regular motions.

\begin{figure}[bt]
\includegraphics{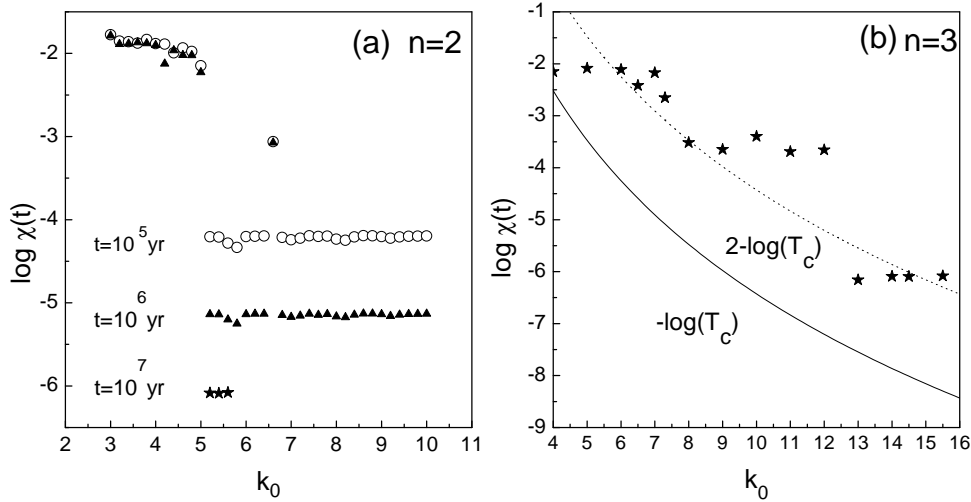}
\vspace{6cm}
\caption{(a) Variations of $\chi(t)$ with $k_0$ at different time:
$10^5$ yr (circles), $10^6$ yr (triangles) and $10^7$ yr (stars) in a 2-planet EMS system
with mass $\mu=10^{-8}$.
(b)Variations of $\chi(t)$ with $k_0$ at $t=10^7$ yr (stars)
 in a 3-planet EMS system with  mass $\mu=10^{-8}$. The solid and dotted curves
plot the functions $-\log(T_c)$ and $2-\log(T_c)$, respectively.
  }
\end{figure}

\section{Conclusion}
In this paper, we study the boundary between chaotic and regular motions in
EMS systems with $n$ planets.   Our results indicate that  the growth  of velocity
dispersion is robust as long as $n\ge 3$,  which gradually turns the orbits
initially in regular regions to chaotic ones. Hence there does not exist
a borderline in the initial separation space, which does not conflict with
the existence of boundaries in the full phase space (position and
velocity space) of N-body systems.

\section*{Acknowledgements}
 We thanks Prof. D.N.C. Lin for helpful discussions.
 This work is supported  by NSFC (No.10233020,10778603) and NCET (04-0468) of China.


\begin{thebibliography}{0}

\bibitem{1} H. Poincar{\'e},, {\it Les M{\'e}thodes Nouvelles de la M{\'e}canique C{\'e}leste} (Gauthier-Villars, Paris, 1892)

\bibitem{2} V. I. Arnold, {\it Dokl. Akad. Nauk SSSR} {\bf 156}, 9 (1964)

\bibitem{3} J. Laskar, {\it Nature} {\bf 338}, 237 (1989)

\bibitem{4} B. Gladman, {\it Icarus} {\bf 106}, 247 (1993)

\bibitem{5} J. Wisdom,  {\it AJ} {\bf 85},1122 (1980)

\bibitem{6} J.L. Zhou, D.N.C. Lin, \& Y.S. Sun, Post-Oligarchic Evolution of  Protoplanetary Embryos and the Stability of
      Planetary Systems, accepted by {\it  ApJ}, arXiv:astro-ph/0705.2164, (2007), (Paper I)

\end{thebibliography}
\end{document}